\documentclass[11pt,a4paper]{article}

\usepackage{amsmath,amsfonts,graphicx,natbib,psfrag,color,enumerate}
\usepackage{algorithm, algorithmic}
\usepackage[textsize=small]{todonotes}
\usepackage{booktabs}

\def\bSig\mathbf{\Sigma}

\renewcommand{\vec}[1]{\boldsymbol{#1}}
\newcommand{\logit}{\text{logit}}
\newcommand{\elogit}{\text{elogit}}

\newcommand{\thetavec}{\vec{\theta}}
\newcommand{\Thetavec}{\vec{\Theta}}
\newcommand{\sigmavec}{\vec{\sigma}}
\newcommand{\rvec}{\vec{r}}
\newcommand{\yvec}{\vec{y}}
\newcommand{\xvec}{\vec{x}}
\renewcommand{\epsilon}{\varepsilon}
\newcommand{\LN}{\text{LN}}

\newcommand{\ag}[1]{\textcolor{black}{#1}}

\graphicspath{{graphics/}}
\usepackage[figuresright]{rotating}

\title{Parameter inference for a stochastic kinetic model of expanded
  polyglutamine proteins}

\author{H. F. Fisher$^{1,2}$,
R. J. Boys$^{1}$, C. S. Gillespie$^{1 *}$\footnote{colin.gillespie@newcastle.ac.uk}, C. J. Proctor$^{3}$, A. Golightly$^{1}$}

\date{\small $^1$ School of Mathematics, Statistics \& Physics, Newcastle University, UK \\
$^{2}$ Population Health Sciences Institute,
Newcastle University, UK \\
$^{3}$ Institute of Cellular Medicine, Newcastle University, UK}

\begin{document}

\maketitle

\begin{abstract}
  The presence of protein aggregates in cells is a known feature of
  many human age-related diseases, such as Huntington's disease.
  Simulations using fixed parameter values in a model of the dynamic
  evolution of expanded polyglutamine (PolyQ) proteins in cells have
  been used to gain a better understanding of the biological system,
  how to focus drug development and how to construct more efficient
  designs of future laboratory-based \emph{in vitro} experiments.
  However, there is considerable uncertainty about the values of some
  of the parameters governing the system. Currently, appropriate
  values are chosen by \textit{ad hoc} attempts to tune the parameters
  so that the model output matches experimental data.  The problem is
  further complicated by the fact that the data only offer a
  partial insight into the underlying biological process: the data
  consist only of the proportions of cell death and of cells with
  inclusion bodies at a few time points, corrupted by measurement
  error.

  Developing inference procedures to estimate the model parameters in this scenario is a significant task. 
  The model probabilities corresponding to the observed proportions cannot be
  evaluated exactly and so they are estimated within the inference algorithm by repeatedly simulating realisations from
  the model. In general such an approach is computationally very expensive and we therefore construct Gaussian process
  emulators for the key quantities and reformulate our algorithm around these fast stochastic approximations. 
  We conclude by examining the fit of our model and highlight appropriate values of the model parameters 
  \ag{leading to new insights into the underlying biological processes such as the kinetics of aggregation.}
\end{abstract}


\section{Introduction}
\label{s:intro}

One of the main aims of modelling biological systems is to describe
and understand the temporal evolution of the system taking account of
the potentially complex inter-relationships between components within
the system. Models can also be used to facilitate \emph{in silico}
experiments in which virtual experiments are performed on a computer.
These \emph{in silico} experiments have an advantage over
laboratory-based experiments as, in general, they are much cheaper and
faster to conduct. This can lead to a better understanding of, for
example, the biological system, how to focus drug development and how
to construct more efficient designs of future laboratory-based
\emph{in vitro} experiments.

The accumulation of abnormal protein deposits within cells are hallmarks of neurodegenerative diseases affecting humans
as they age. \ag{There are many such diseases, (e.g. Alzheimer’s and Parkinson’s disease), in which different parts of
  the brain are affected resulting in a range of clinical symptoms such as loss of motor function, dementia, and
  behavioural changes. Despite differences between symptoms, there are similarities in the underlying molecular
  mechanisms leading to the accumulation of protein aggregates and the neuronal cell death (Gan, Cookson et al. 2018).
  Although, age is the greatest risk factor for neurodegeneration, there is a group of diseases that are caused by
  genetic mutations. In nine of these diseases, the mutation occurs in a gene that contains a repeat of a CAG nucleotide
  triplet (Lieberman, Shakkottai et al. 2019). As CAG encodes for the amino acid glutamine, these diseases are known as
  polyglutamine diseases and the proteins they encode are referred to as PolyQ proteins, although different genes and
  proteins are involved in each disease, e.g. Huntington’s disease (HD), (Lieberman, Shakkottai et al. 2019). HD is an
  adult-onset, progressive disease characterised by loss of motor function, psychiatric disorders and dementia (McColgan
  and Tabrizi 2018). In this disease, the CAG triplet is found in the huntingtin gene (HTT, HGNC:4581). In the normal
  HTT gene, the CAG triplet is repeated 10-35 times but in the mutated form, the segment is expanded from 36 to over 120
  repeats. This leads to the formation of an abnormally long protein which may be cleaved within the cell into small
  toxic fragments that enter the nucleus where they form aggregates, known as inclusion bodies (or inclusions), and
  sequester other proteins causing transcriptional dysregulation although the exact mechanisms of how this occurs are
  still unknown (McColgan and Tabrizi 2018). In addition, huntingtin fragments form aggregates in the cytoplasm and may
  impair cellular function, e.g. impairment of proteostasis (McColgan and Tabrizi 2018).}

\ag{There are currently no effective interventions for the prevention, delay in onset or slowing down of disease
  progression in HD or in any other neurodegenerative disorder. This is mainly due to a lack of a full understanding of
  the underlying molecular mechanisms. In particular, although protein aggregation is a common feature of all these
  disorders, it is still not fully known how protein aggregates actually contribute to the disease process. There has
  been considerable controversy over which stage of the aggregation process is most toxic to cells, and it has been
  suggested that the formation of large inclusion bodies may be protective as they sequester misfolded proteins and
  prevent the overload of protein degradation pathways (Ross and Poirier 2004). This has been shown experimentally in
  cell culture (Arrasate, Mitra et al. 2004). However, this protective effect may only be short-term, as large
  aggregates also contain proteins involved in the removal and repair of damaged protein and it has been suggested that
  they may induce quiescence and induction of cell death via necrosis (Ramdzan, Trubetskov et al. 2017), which can cause
  damage to neighbouring cells. In addition, nuclear inclusions may impair redox signalling leading to an increase in
  oxidative stress and further damage (Paul and Snyder 2019). The controversy regarding cytoxicity of PolyQ proteins is
  largely due to insufficient understanding of the molecular mechanisms involved. This motivated our previous study
  which used live cell imaging with fluorescent reporter systems to examine the relationship between PolyQ protein,
  activation of the stress kinase p38MAPK (MAPK14; HGNC:6876 ), reactive oxygen species (ROS) generation, inhibition of
  the proteasome (a protein complex which degrades cellular proteins), and formation of PolyQ nuclear inclusions (Tang,
  Proctor et al. 2010). It was found that proteasome inhibition usually preceded formation of inclusion bodies and that
  p38MAPK inhibitors alleviated the inhibition of proteasomes and delayed the onset of inclusion formation. Conversely,
  the addition of proteasome inhibitors resulted in earlier formation of inclusions. This study also included a
  computational model which explored }the complex interactions of PolyQ proteins with other elements of the cell and
they used computer simulations from the model (with fixed parameter values) to suggest ways to reduce the toxicity of
PolyQ proteins on cells. As their model describes dynamic interactions at the single cell level, the number of different
biochemical species vary discretely over time, often with low copy numbers~\citep{Gill77}. Also the interaction of the
species is driven by Brownian motion and so accurate modelling requires that it take account of the inherent
stochasticity present in the system.

Currently, plausible values for the parameters in the PolyQ model are
determined by using model simulations for specified values of the
parameters and then adjusting them in an attempt to match experimental
data. Clearly this is a difficult task and one with which
statisticians can make a contribution. The general scenario of
modelling biological systems in computational systems biology is
described by \cite{kitano2001} with increasing focus on those
describing the stochastic dynamics of individual cells
\citep{ElowitzLSS2002, SwainES2002}. More recently an overview of this
area has been provided by \cite{VillaverdeB2014}.  

A key difficulty in conducting parameter inference for \ag{complex 
stochastic models such as the PolyQ model} is that the experimental data 
\ag{are typically incomplete and subject to measurement error. This necessitates 
the use of computationally intensive schemes such as Gibbs sampling 
\citep{BoysWK08}, pseudo-marginal Metropolis-Hastings 
\citep{golightly11,Georgoulas2016,wilkinson2018stochastic}, population 
Monte Carlo \citep{koblents2015,koblents2019bayesian} and approximate 
Bayesian computation \citep{Wu2014,owen2015}. Inference is particularly 
challenging for the PolyQ model since the data consist only of the 
proportion of cells which are dead and the proportion of cells 
which contain inclusion bodies. Since the corresponding model probabilities 
are intractable, these latent proportions must be estimated at each 
observation time by repeatedly forward simulating the model over 
the duration of the time-course. Computational cost can be reduced 
by replacing the expensive and exact stochastic simulator with a cheap 
approximation. For example, approaches which replace the continuous-time stochastic 
model with a discrete-time approximation include tau-leaping 
\citep{Gillespie2001,Cao2006} and the chemical Langevin equation \citep{Gillespie00,kampen2001}.} 


\subsection{Contributions and organisation of the paper}

The aim of this paper is to examine what insight the limited available
experimental data provides about the PolyQ model parameters and to check
whether the model gives a reasonable description of the dynamic
variation observed in the experimental data.

\ag{Given the prohibitive computational cost associated with fitting the stochastic polyQ model, 
we consider a computationally feasible approximation and perform exact (simulation-based) 
inference for the resulting model. We eschew the use of an approximate simulator in favour of \emph{directly emulating} 
the empirical logit of the proportions of interest with a Gaussian 
process (GP) \citep{Rasmussen2006,SantnerWN03}. In particular, combining the 
fitted GP emulator with a Gaussian measurement model (on the same scale 
as the observed data) allows a direct approximation of the observed data likelihood 
without recourse to further simulation from the stochastic PolyQ model.}

\ag{The findings from this analysis give new information regarding the parameters of the
model, which in turn leads to new insights into the underlying biological processes. For
example, this analysis has shown that during the first stages of the aggregation process both
aggregation and disaggregation probably occur much faster and that the
disaggregation/aggregation ratio is likely to be a magnitude higher than was originally
assumed. The biological implication of this is that it will take longer to reach the threshold
size required for inclusion formation but that there will also be higher levels of small
aggregates which will inhibit the proteasome.}

\ag{The remainder of the paper is organised as follows.} 
Section~\ref{sec:exptdata} details the experimental data available and
outlines how it was collected. A complete description of the 
model is given in section~\ref{sec:stochmodel} which includes 
the underlying stochastic model which
describes the dynamic evolution of this biological process and the 
observation model, which links the data to the underlying 
process. Our prior assumptions about parameter 
values and initial levels are given in~\ref{sec:prior}. A method for
determining the posterior distribution for model quantities is
described in section~\ref{sec:posterior} and, because a standard
simulation-based MCMC solution is prohibitively expensive, we develop
Gaussian process emulators in section~\ref{sec:emulation} which
facilitate timely generation of realisations from the posterior
distribution. As we have high prior uncertainty on model parameters,
in section~\ref{sec:historymatching} we employ a history matching
technique which removes implausible training points in an attempt to
fit the GPs within regions of non-negligible posterior support. We also 
include an
assessment of the quality of the final fitted GPs. We present our findings about
the the PolyQ model in section~\ref{sec:results}, and point out new
insights into this biological system.

\section{Experimental data}
\label{sec:exptdata}
We have data from two different sets of experiments. These were
carried out in the same laboratory and are described briefly below. A
more comprehensive description of the experimental procedures can be
found in~\cite{Tang2010}.

\subsection{Cell death}
The first experiment begins with a large number of human U87MG
glioblastoma cells maintained in a suitable medium at 37$^{\circ}$C at
5\% CO$_2$. The cells are transfected with a construct that contains
the expanded PolyQ Huntington protein (see section~\ref{sec:ibf} for
details). The survival of cells is monitored over time. Three
different repeats of the experiment are carried out under the
following experimental conditions:
\begin{enumerate}[(i)]
\item \emph{Control group}: No intervention.
\item \emph{Proteasome inhibition group}: Cells are treated with a
  proteasome inhibitor 24 hours after transfection. The stochastic
  kinetic model captures this scenario by reducing the initial value
  $k_{proteff}=1$ to $k_{proteff}=0.05$ after 24 hours. 
\item \emph{p38 inhibition group}: Cells were pre-treated for 2 hours
  with a p38 inhibitor. The stochastic kinetic model captures this
  scenario by setting $k_{p38act}=0.05$. 
\end{enumerate}
The changes to parameter values used to describe the different
experimental conditions were determined by conversations with the
experimentalists. It was felt that reducing the rates to zero in
conditions (ii) and (iii) overstated the effect of the experimental
treatment and so these rates are set to low but non-zero values.

The cells are stained with propidium iodide, which is a fluorescent
dye with the property that it only binds to non-viable (dead) cells.
This technique is called \emph{propidium iodide exclusion} and is used
to identify the viability of the cells over time. The fluorescent dye
can be viewed under a microscope and estimates of the proportion of
cell death can be observed over time. These proportions of cell death
form our experimental data which can be seen in
Table~\ref{tab:polyq_data}.  Each row of the table corresponds to
experiments carried out under the different experimental conditions
outlined above. We have two repeats of each experiment.

Within the stochastic model, cell death can occur via two biological
pathways, either via proteasome inhibition or by p38 activation. This
process is monitored within the model by using the dummy species
\texttt{PIdeath} and \texttt{p38death}. These species are both binary
variables, with a zero value representing that the cell is alive.

\begin{table}
  \caption[PolyQ model: experimental data]{Proportions of cell death
    observed under different experimental conditions. }
  \label{tab:polyq_data}
  \centering
  \begin{tabular}{@{}llll@{}} 
\toprule
Condition & 24hrs & 36hrs &  48hrs \\ 
\midrule
(i) & 0.1503 & 0.1455 & 0.2608 \\ 
& 0.1788 & 0.1821 & 0.2846 \\
(ii) &  0.1897 & 0.1807 & 0.2250 \\
& 0.1640 & 0.1973 & 0.2998 \\ 
(iii) & 0.2168 & 0.2344 & 0.3644 \\ 
& 0.2436 & 0.2095 & 0.3872 \\
\bottomrule
  \end{tabular}
\bigskip
\end{table}

\subsection{Inclusion body formation}
\label{sec:ibf}
Experimentalists monitor the number of cells with inclusion bodies by
first creating a construct that encodes an expanded PolyQ Huntington
protein which is tagged with Yellow Fluorescent Protein (YFP). U87MG
cells are transfected with the construct and then imaged at 10 minute
intervals between 24 to 48 hours after transfection. Time lapse images
reveal the formation of inclusions (by detecting YFP) and the
percentage of cells with inclusions is recorded every 6 hours. The
experimental results are given in
Table~\ref{tab:polyq_inclusion_data}. As with the data on cell death,
each row corresponds to experiments carried out under different
experimental conditions and we have two repeats of each experiment.

Within the stochastic model, the number of cells containing inclusions
is counted via the number of cells containing \texttt{SeqAggP}. In the
laboratory, image analysis suffers from a thresholding problem in
detecting inclusions and we capture this aspect by defining cells with
inclusions as those in which \texttt{SeqAggP} contains more than ten PolyQ
proteins (i.e. $\texttt{SeqAggP}>10$). The value of this threshold is
based on the biological modeller's best understanding of the
mechanism, though in fact the actual value is not crucial since
aggregates grow very rapidly and there is only a very small timeframe
when \texttt{SeqAggP} is less than ten.
\begin{table}
  \caption[PolyQ model: experimental data]{Proportions of inclusion
    bodies observed under different experimental conditions.}
  \label{tab:polyq_inclusion_data}
  \centering
  \begin{tabular}{@{}l lllll@{}} 
\toprule
Condition & 24hrs & 30hrs & 36hrs & 42hrs & 48 hrs \\ 
\midrule
(i) &  0.0909 & 0.2857 & 0.3824 & 0.4286 & 0.7273 \\ 
 &  0.0175 & 0.2131 & 0.3538 & 0.4304 & 0.5742 \\ 
(ii) &  0.0909 & 0.4247 & 0.6125 & 0.7403 & 0.8194 \\
 &  0.1154 & 0.5075 & 0.7895 & 0.8667 & 0.9157 \\ 
(iii) &  0.0303 & 0.0444 & 0.0612 & 0.1373 & 0.2200 \\ 
 &  0.0189 & 0.0938 & 0.1286 & 0.1613 & 0.1833 \\ 
\bottomrule
  \end{tabular}
 \bigskip
\end{table}

\section{The stochastic model}
\label{sec:stochmodel}
\subsection{PolyQ mechanism}
\label{sec:polyQmechanism}
\begin{figure}
	\centering
	\includegraphics[width=0.5\textwidth]{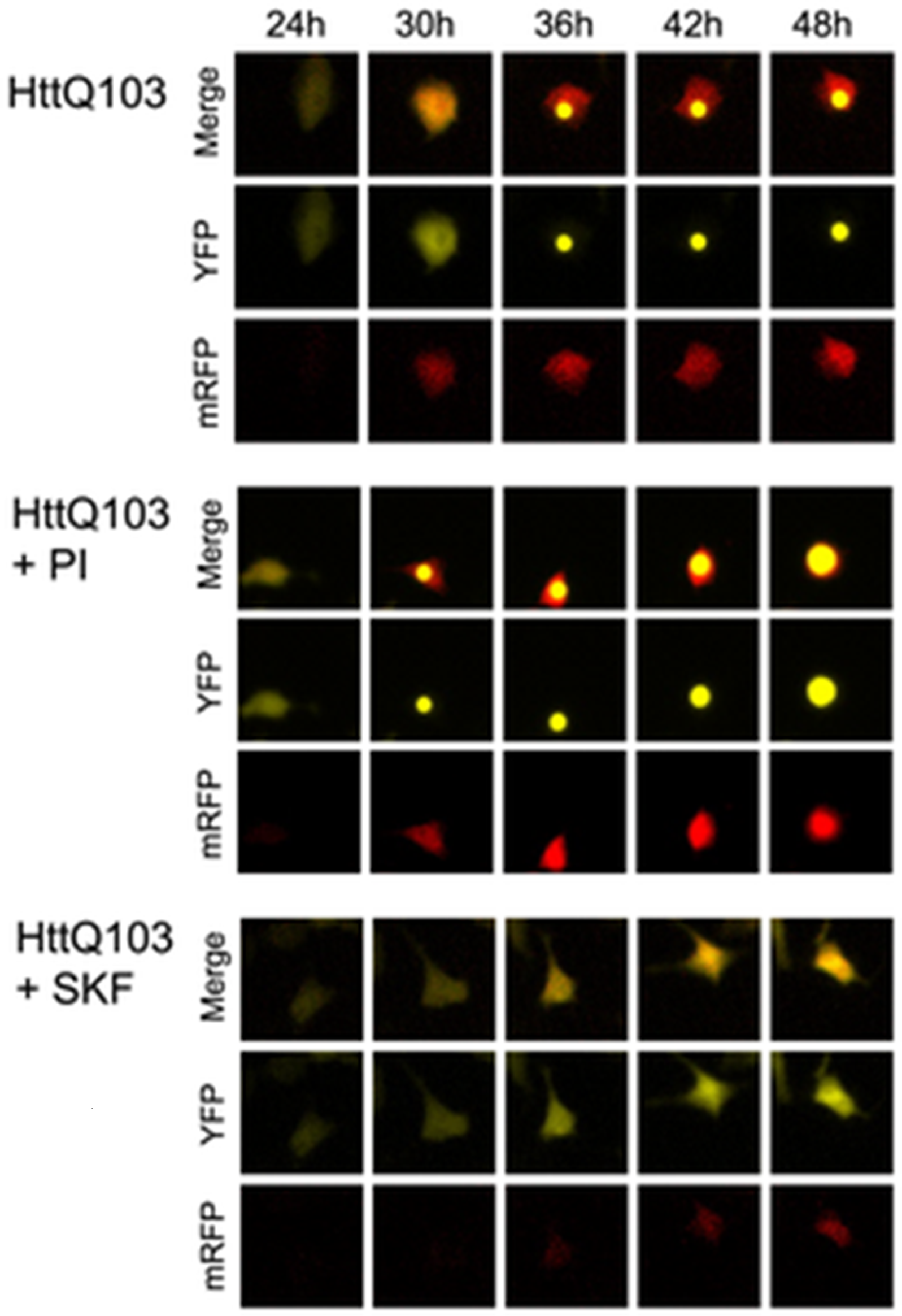}
	\caption{\ag{Example data from live cell imaging.}}
	\label{fig:polyq_experiment}
\end{figure}

The stochastic model for the PolyQ mechanism we consider in this paper
is a reduced form of that proposed by~\cite{Tang2010}. \ag{The original model developed by Tang et al. (Tang, Proctor et al. 2010) was
constructed in the Systems Biology Markup Language (SBML) and is available from the Biomodels
database (Li, Donizelli et al. 2010) (Model ID: BIOMD0000000285). The model was developed to
investigate the effects of PolyQ on proteasome function, oxidative stress, and cell death. It
contained a pool of PolyQ which was continually turned over, being degraded by the proteasome. It
was assumed that the PolyQ aggregation process started by two monomers interacting, which then
interacts with further monomers causing the aggregate to grow (\texttt{AggPolyQ1-5}). At early stages, it
was assumed that disaggregation could also take place but that when the aggregate reached a
threshold size (denoted by \texttt{SeqAggP} in the model), disaggregation does not take place, so that the
aggregate continues to grow and an inclusion forms. The model also included a generic pool of
protein which could be damaged by ROS leading to misfolding and aggregation. It was assumed that
small aggregates bind to the proteasome but cannot be degraded and so inhibit proteasome
function. In addition, they may increases levels of ROS. The model also included the stress kinase
p38MAPK (simply modelled as two pools: inactive [\texttt{p38}] and active [\texttt{p38P}]) with activation occurring
more frequently when levels of ROS are high (Sato, Okada et al. 2014), and that high levels of \texttt{p38P}
activated a cell death pathway (denoted by \texttt{p38death}). Proteasomes bound by aggregates
(\texttt{AggPProteasome}) could also activate a cell death pathway (denoted by \texttt{PIdeath}). The model also
included the turnover of a fluorescent protein (\texttt{mRFPu}) as an increase in \texttt{mRFPu} levels indicates that
the proteasome is inhibited, shown by live cell imaging in Tang et al. 2010. An example of this data is
shown in Figure\ref{fig:polyq_experiment}.}

\ag{Given the complexity of the original model (with 27 chemical species and 72 reactions) 
it was necessary to consider a simplification for 
this study.} The changes to the Tang model have been to reduce the number of species and reactions by removing the
generic pool of protein and its associated reactions of misfolding and aggregation, and by removing the fluorescent
protein mRFPu, which had been originally included as a marker of proteasome function. \ag{In the process of 
model construction, it was found neccesary to add in
an effect of ROS on the aggregation process. The experimental
biologists suggested that ROS increases the propensity of PolyQ
proteins to aggregate due to ROS interfering with the
ubiquitin-proteasome system (UPS), so that PolyQ proteins are more
likely to aggregate than be degraded when ROS levels are high. The
exact mechanism is not fully understood and so we simply included ROS
in the rate laws for PolyQ aggregation.}

The resulting reduced model 
contains 14 chemical species and 33 reaction channels, which is
still relatively large when compared to other stochastic kinetic models for which fully Bayesian inference is available
in the literature \citep{BoysWK08,HendersonBPW10}. 
A complete list of the reactions and their stochastic rate laws describing the PolyQ
model can be found in Appendix~A of the Supplementary Materials. All reactions
except those relating to aggregation follow the law of mass action
kinetics. \ag{We used a Hill function for
the effects of ROS as it is thought that low levels of ROS would have
little effect on the UPS with a maximal effect when ROS levels are
high. We assumed that the rate of aggregation would be half the
maximal rate when ROS is at its basal level, which is achieved by
using the value 10 in the denominator of the rate law.}

\begin{figure}
\includegraphics[width=\textwidth]{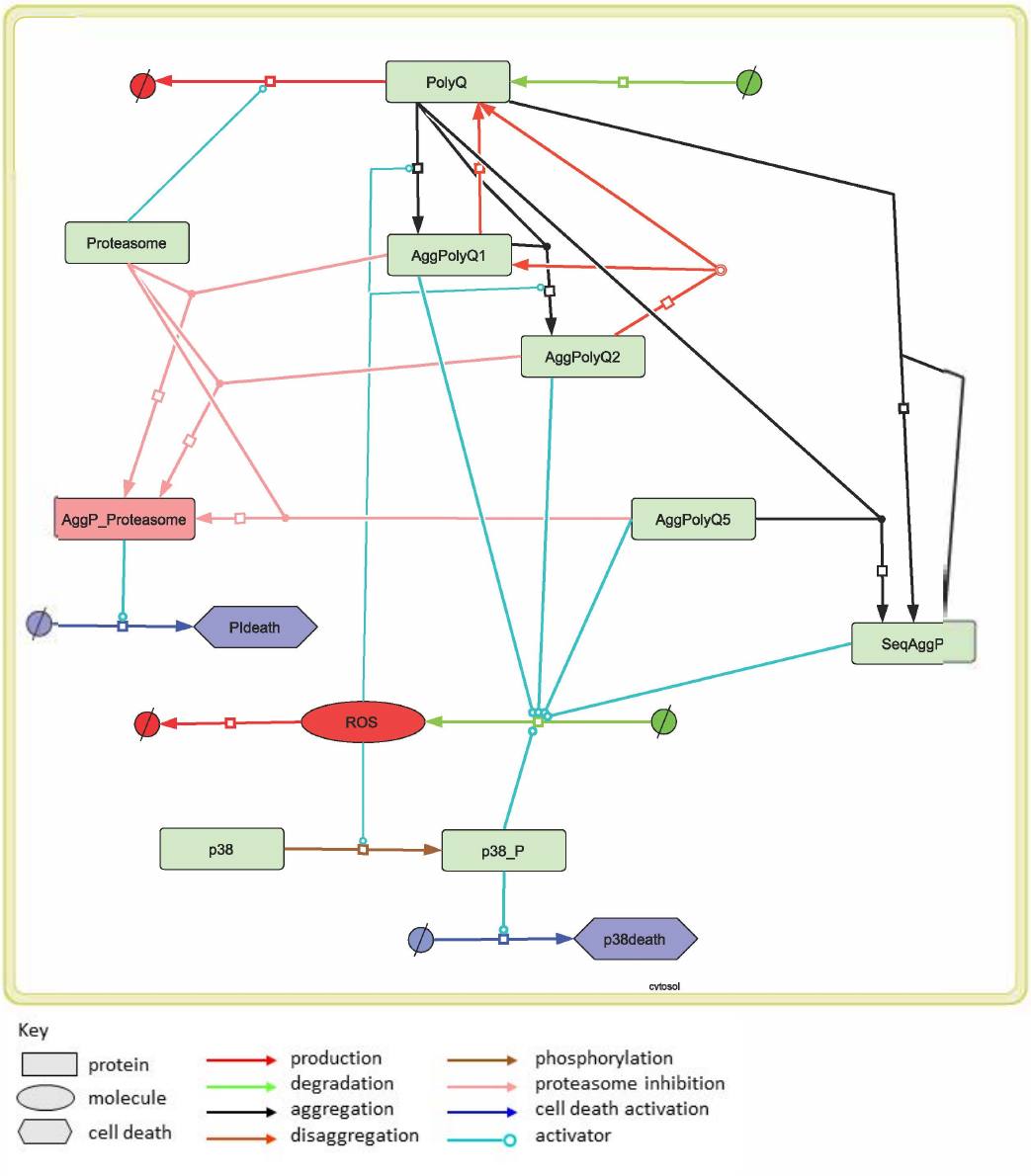}
\caption{\label{fig:polyq}Network diagram of the PolyQ model}
\end{figure}

The model is represented graphically in Figure~\ref{fig:polyq}, where oval shapes (nodes)
represent chemical species and an arrow between two nodes represents a reaction which can take place involving the two
chemical species. The figure clearly shows the complex interlinkages and complex feedback loops within the model.

\subsection{The observation model}
\label{sec:obsnmodel}
We assume that the experimental data are noisy versions of the
equivalent quantities in the stochastic model and adopt a simple model
for this observation error which is additive independent normal noise
on a logit scale (as the data are proportions). Other error models are
possible, such as embedding the model probabilities within a Beta
distribution, but this logistic normal model affords advantages in the
subsequent analysis.

Suppose the stochastic model for the experiment run under condition
$c$ with parameter vector $\thetavec$ gives $p_{t,c}^D(\thetavec)$ and
$p_{t,c}^I(\thetavec)$ as the probabilities of death and of the
presence of inclusions at time $t$ and the logits of their observed
versions as $y^D_{t,c,r}$ and $y^I_{t,c,r}$ in experimental
repeat~$r$.  We assume the observation model is, for $c=1,2,3$,
$r=1,2$
\begin{align*}
y^D_{t_i,c,r}
&=\logit\,p_{t_i,c}^D(\vec{\theta})+\sigma_D\epsilon_{i,c,r},
\quad i=1,2,3\\
y^I_{t_i,c,r}
&=\logit\,p_{t_i,c}^I(\thetavec)+\sigma_I\epsilon_{i,c,r}^*, 
\quad i=1,\ldots,5
\end{align*}
where the $\epsilon_{i,c,r}$ and $\epsilon_{i,c,r}^*$ are i.i.d.
standard normal quantities and $t_i$ denote the $i$th time-point at 
which an observation is available.  Note that this model assumes the same
measurement error distribution for each experimental condition and
repeat. The implications of this observation model on the probability
scale is explored in Appendix~B of the Supplementary
Materials. 

\section{Prior information}
\label{sec:prior}
Tables~\ref{tab:polyq_parameters_fixed} and \ref{tab:polyq_parameters} contain a list of all parameters in
the model. Some parameters are known quite accurately in the
literature. For example, the synthesis and degradation rates of PolyQ
proteins $k_{synPolyQ}$ and $k_{degPolyQ}$ can be obtained by
considering the half-life of PolyQ proteins~\citep{Persichetti1996}.

\begin{table}[t]
  \caption{Parameters in the PolyQ model under condition (i) with their known
    values}\label{tab:polyq_parameters_fixed} 
  \centering
  \begin{tabular}{@{} llll @{}}
    \toprule
    Parameter & Value & Units & Reference\\
    \midrule
    $k_{synPolyQ}$ & $0.25$ & molecule $s^{-1}$ & Calculated from value of $k_{degPolyQ}$ to give \\
    & & & steady state level of 1000 molecules\\
    $k_{degPolyQ}$  & $2.5 \times 10^{-7}$  &  molecule$^{-1} s^{-1}$ & Half-life of $30$ hours (Persichetti et al., 1996) \\
    $k_{actp38}$ & $2 \times 10^{-3}$ & molecule$^{-1} s^{-1}$ &  Phosphorylation occurs within \\
    & & & (Aoki,Takahashi et al. 2013)\\
    $k_{p38act}$  & 1.0 & dimensionless & Dummy variable (set to $0.05$ for p38 inhibition)\\
    $k_{proteff}$ & 1.0 & dimensionless & Dummy variable (set to $0.05$ for proteasome \\
                                     & & & inhibition)\\
    \bottomrule
  \end{tabular}
\end{table}

\begin{table}
  \centering
  \caption[PolyQ model: parameters and values used for simulating
  data]{Parameters in the PolyQ model under condition (i) with their prior distribution.}
  \label{tab:polyq_parameters} 
  \begin{tabular}{@{}ll @{}}
    \toprule
Parameter  & Prior distributions\\
\midrule
$k_{aggPolyQ}$ & $\theta_1=\log k_{aggPolyQ}\sim N(\log 10^{-7},5)$\\
$k_{disaggPolyQ_1}$ & $\theta_2 = \log k_{disaggPolyQ_1} \sim N(\log(5\times 10^{-7}),5)$\\
$k_{seqPolyQ}$ &  $\theta_3 =\log k_{seqPolyQ} \sim N(\log(8\times 10^{-7}),5)$ \\
$k_{inhprot}$ & $\theta_4=\log k_{inhprot}\sim N(\log(5\times 10^{-9}),5)$\\
$k_{remROS}$ & $\theta_5 = \log k_{remROS} \sim N(\log(2\times 10^{-4}),5)$\\
$k_{genROSSeqAggP}$ & $\theta_6 =\log k_{genROSSeqAggP} \sim N(\log(10^{-7}),5)$\\
$k_{genROSAggP}$ & $\theta_7=\log k_{genROSAggP}\sim N(\log(5\times 10^{-6}),5)$\\
$k_{inactp38}$ & $\theta_8 = \log k_{inactp38} \sim N(\log 0.8,5)$\\
$k_{genROSp38}$ & $\theta_9 =\log k_{genROSp38}\sim N(\log(7\times 10^{-7}),5)$\\
$k_{p38death}$ & $\theta_{10}=\log k_{p38death}\sim N(\log(9\times 10^{-8}),5)$\\
$k_{PIdeath}$ & $\theta_{11}=\log k_{PIdeath}\sim N(\log(2.5\times 10^{-8}),5)$\\
$\sigma_D$ & $\sigma_D\sim InvChi(2,0.12)$\\
$\sigma_I$ & $\sigma_I\sim InvChi(0.75,0.05)$ \\
\bottomrule
\end{tabular}
\bigskip
\end{table}

Also, to reduce the number of parameters in the model, it was thought
reasonable to fix some rates to be identical to others or to be
functions of other rates. For example, the disaggregation rates for
different sized PolyQ aggregates have been fixed as known proportions
of the rate for single PolyQ aggregates:
$k_{disaggPolyQ2}=0.8\,k_{disaggPolyQ1}$,
$k_{disaggPolyQ3}=0.6\,k_{disaggPolyQ1}$, and so on. In each case,
these identities or fixed proportions have been chosen according to
the biological modeller's best understanding of the reaction system.

Only limited information is available for the remaining stochastic
rate parameters and we represent these fairly weak prior beliefs using
independent log-normal distributions with medians set at our
biological modeller's best guess. On the log scale, the kinetic rate
parameters are denoted by $\theta_1,\ldots,\theta_{11}$ and these
parameters have independent normal prior distributions. The weak
beliefs are represented by prior variances of five on the log scale as
these correspond to a plausible range of values for the
(untransformed) kinetic rates from 0.01 to 100 times their median
value.

We have prior beliefs for the level of measurement error in the cell
death experiments ($\sigma_D$) from time course data on two
independent replicates from an independent (but similar) study
measuring cell death proportions. The difference between these
replicates on the logit scale form a random sample from a
$N(0,\sigma_D^2)$ distribution and so these logit differences lead to
an inverse gamma posterior for $\sigma_D^2$ under a vague prior. This
posterior corresponds to an inverse chi distribution for $\sigma_D$,
where this distribution is defined as that of $1/\sqrt{X}$, where $X$
has a gamma distribution. Further, as these data arise from a similar
but different study, we choose to construct our prior as a powered
down version of this inverse chi posterior as this makes it more
diffuse around an appropriate value \citep{IbrahimC2000}. Our
knowledge about the level of measurement error in the inclusion body
experiments is weaker and so we use an even more powered down prior
for~$\sigma_I$.

There is also uncertainty about some of the initial levels of the
different species. This uncertainty is captured by independent prior
distributions elicited from the biological modeller. Other specie
levels are known due to the construction of the experiment, such as
those for \texttt{AggPolyQ$_1$}--\texttt{AggPolyQ$_5$}, others obey a
conservation law ($\texttt{p38}+\texttt{p38P}=100$) and some levels,
such as that for this conservation level and for \texttt{PolyQ}, set
at fairly arbitrary high values reflecting the relative abundance of
these species within the cell. The values or prior distribution of the
initial levels are listed in Table~\ref{tab:polyq_species_short}. Note
that two of these initial levels have a discrete log-normal $DLN$
prior distribution, with probability function
$\pi(i)=Pr(i-0.5<X<i+0.5)$, where $X$ is a log-normal random variable.

\begin{table}
\centering
\caption[PolyQ model: Value or prior distribution of the initial
  levels of the chemical species]{\label{tab:polyq_species_short}
  Value or prior distribution of the initial levels of the chemical
  species}
\begin{tabular}{@{}ll@{}}
\toprule
Name & Value or prior distribution\\
\midrule
\texttt{PolyQ}  & 1000 \\ 
\texttt{Proteasome}  & $DLN(6.9,0.1^2)$\\ 
\texttt{AggPolyQ$_{1-5}$}  & 0 \\
\texttt{SeqAggP} & 0 \\
\texttt{AggPProteasome}  & 0 \\
\texttt{ROS}  & $DLN(2.5,0.25^2)$\\ 
\texttt{p38P} & $U\{0,1,\ldots,5\}$ \\
\texttt{p38}  & 100-\texttt{p38P} \\
\texttt{PIdeath} & 0 \\
\texttt{p38death}  & 0 \\
\bottomrule
\end{tabular}
\bigskip
\end{table}

\section{Posterior inference}
\label{sec:posterior}
Knowledge of the kinetic rates $\thetavec$ and the observation error
levels $\sigmavec=(\sigma_D,\sigma_I)$ is summarised in the posterior
distribution, which has density
\[
\pi(\thetavec,\sigmavec|\yvec)
\propto\pi(\thetavec)\pi(\sigmavec)\pi(\yvec|\thetavec,\sigmavec),
\]
where $\yvec$ represents the collection of all data on cell death and
inclusions. Although the normal likelihood has a simple form, it does
require calculation of the probabilities $p_{t,c}^D(\thetavec)$ and
$p_{t,c}^I(\thetavec)$. Unfortunately analytic expressions for these
probabilities are not available due to the complexity of the
stochastic model. Prior uncertainty around the initial specie
levels~$\xvec_0$ is a further complicating factor as, for example,
$p_{t,c}^D(\thetavec)=E_{\xvec_0}\{p_{t,c}^D(\thetavec,\xvec_0)\}$, 
where $E_{\xvec_0}\{\cdot\}$ denotes expectation with respect to 
$\xvec_0$. 
Therefore we estimate these probabilties using $n$ independent
realisations of the stochastic model, initialised according to the
prior distribution on initial levels. Suppose that a typical
probability is estimated by the proportion $\widehat{p}_n$. For
sufficiently large $n$, the empirical logit of such proportions
$\elogit\,\widehat{p}_n=\log\{(\widehat{p}_n+0.5/n)/(1-\widehat{p}_n+0.5/n)\}$
is unbiased for $\logit~p$ and its sampling variability is closely
described by a normal distribution with variance
$1/\{n\widehat{p}_n(1-\widehat{p}_n)\}$. Thus, taking an improper
constant prior for $\logit~p$ gives its posterior distribution as a
normal distribution with mean $\elogit\,\widehat{p}_n$ and variance
$1/\{n\widehat{p}_n(1-\widehat{p}_n)\}$. Therefore we can integrate
out posterior uncertainty about $\logit~p$ in the observation model,
modifying it to have independent components of the form
\[
y\sim N\bigl(\elogit\,\widehat{p}_n(\thetavec),~
\sigma^2+1/[n\,\widehat{p}_n(\thetavec)\{1-\widehat{p}_n(\thetavec)\}]\bigr).
\]
Building an MCMC inference scheme around this normal likelihood is
straightforward. We found that estimating the probabilities using
$n=1000$ model simulations gave empirical logits which fitted well to
a normal distribution. Note that, in order for these estimates to be
uncorrelated, they have each been calculated using independent
realisations of the model. 

Unfortunately the MCMC scheme requires the generation of many millions
(or even billions) of realisations from the stochastic model as it
explores the posterior distribution.  Clearly this problem prohibits
using such a scheme for obtaining realisations from the posterior
distribution in a timely manner.  In such situations it is commonplace
to use stochastic approximations for these deterministic model
probabilities.  We will use Gaussian process (GP) approximations
(sometimes called emulators) which are popular in the deterministic
computer model literature and elsewhere.  Useful background articles
in this area are \cite{SantnerWN03} and \cite{BayarriEtal07}. The
suitability of GPs as emulators is highlighted in
\cite{O'Hagan06}. Also see \cite{HendersonBKLW} for an illustration of
the utility of GP emulators in the analysis of complex biological
models.

\section{Emulation}
\label{sec:emulation}
We need to build GP emulators approximating the stochastic output of
the proportion of cell death and that of inclusions under the three
different conditions.  Time ($t$) is an input variable in these six
emulators, in addition to the unknown stochastic rate constants
($\thetavec$) and observation error levels ($\sigmavec$). However, it
can be problematic to specify appropriate covariance kernels over
time. In any case, for Bayesian inference, all that is needed are
emulators for the distributions of the probabilities under each
experimental condition at the distinct time points ocurring in the
datasets. Therefore we will build a total of 24 emulators: 9 for
proportions of cell death and 15 for proportions of inclusion
bodies. Using these time-condition-specific emulators also has the
advantage that they can be built in parallel. Thus we build GP
emulators for
\begin{align*}
x_{t_i,c}^D(\thetavec)&=\elogit\,\widehat{p}_{t_i,c}^D(\thetavec) \quad i=1,2,3
\intertext{and}
x_{t_i,c}^I(\thetavec)&=\elogit\,\widehat{p}_{t_i,c}^I(\thetavec) \quad i=1,\ldots,5,
\end{align*}
for each condition $c=1,2,3$. The aim is to replace the empirical
proportions in the observation model with GP emulators, allowing for
their uncertainty. Fortunately this is straightfoward as each emulator
prediction is also normally distributed and so, as before, the
additional uncertainty introduced by the GPs can be integrated out
analytically and there is no need for the MCMC scheme to integrate
over an additional latent layer within the model.

\subsection{Training data}
In order to build the emulators for the $x_{t_i,c}^D(\thetavec)$ and
$x_{t_i,c}^I(\thetavec)$, we first need to evalute their values at
some set of chosen values for $\thetavec$.  This amounts to choosing
the size and values in an $n_d$-point design
$\Thetavec=(\thetavec_1,\ldots,\thetavec_{n_d})$. In common with other
work in this area \citep{SantnerWN03,HendersonBPW10} we choose to base
our design on a Latin hypercube design (LHD) constructed using the
\emph{maximin} algorithm of~\cite{Morris1995}.  These designs are
popular as they produce an effective and efficient coverage of a
bounded space. We take the bounded space to be the central hypercube
of the prior distribution defined by the marginal 99\% intervals for
the rates~$\theta_j$.  Inevitably, determining an appropriate design
from which to contruct GP emulators is a sequential process.  This is
because it is not unusual to find that there are only a few design
points in the region of high posterior support, particularly when the
marginal priors have large uncertainty (as we have here). To help with
this problem, as part of the sequential building of the emulators, we
filter out design points which are implausible (inconsistent with the
data) by using a history matching technique (see
section~\ref{sec:historymatching}). Note that obtaining estimates at
each point in the LHD for the proportions of cell death and of
inclusions over repeated simulations of the stochastic model is easily
parallelisable on a high performance computer system.

\subsection{Mean function and covariance function}
\label{sec:mean_and_cov}
The mean function was taken as a linear predictor in the components 
of $\thetavec$, 
with the least squares estimates as the coefficients. Thus each
emulator has a mean function at input
$\thetavec=(\theta_1,\ldots\theta_{11})^T$ of the form
\[
m(\thetavec)
=\hat\beta_0+\sum_{i=1}^{11}\hat\beta_i\theta_i.
\]
This choice was taken to give a parsimonious yet reasonably well
fitting mean function and essentially leaves the residuals to this fit
being modelled by a zero mean function Gaussian process. \ag{As is commonly 
used in the emulation literature, we specified a squared exponential 
covariance function, with}
\[
K(\thetavec_i,\thetavec_j|a,\rvec)
=a\exp\left\{-\sum_{k=1}^{11}r_k^2(\theta_{ik}-\theta_{jk})^2\right\}
\]
\ag{where, for example, $\theta_{ik}$ denotes the $k$th component 
of $\thetavec_i$. The parameters of this function control the overall 
level of variability and smoothness of the process, with smaller values 
of the inverse length scales $r_k$ giving smoother realisations.}

\subsection{Hyperparameter estimation}
The hyperparameters $a$ and $\rvec=(r_i)$ for each emulator need to be
estimated from the training data before we can use them as part of the
MCMC inference scheme. When fitting a typical GP to training data
$\xvec(\vec\Theta)=(x(\thetavec_1),\ldots,x(\thetavec_{n_d}))^T$, the
likelihood for the hyperparameters results from
$\xvec(\vec\Theta)|a,\rvec\sim
N_{n_d}(\vec{m}(\Thetavec),K(\Thetavec,\Thetavec|a,\rvec))$ where 
$N_{n_d}(\cdot,\cdot)$ denotes an $n_d$-dimensional Gaussian distribution and 
$K(\Thetavec,\Thetavec|a,\rvec)_{ij}=
K(\thetavec_i,\thetavec_j|a,\rvec)$. In general it is not possible to
obtain a posterior distribution in closed form for a prior on
$(a,\vec{r})$. Here we take independent weak log-normal prior
distributions for the hyperparameters, with $a\sim\LN(0,100)$ and
$r_i\sim\LN(0,100)$, $i=1,\ldots,11$. The GPs were fitted by using the
`no-U-Turn' Hamiltonian Monte Carlo sampler and implemented via the
RStan package interface to the Stan package~\citep{stan,rstan}.

\subsection{Modified observation model}
For a typical GP, the prediction at a point $\thetavec^\dagger$ has
distribution
\begin{equation}
x(\thetavec^\dagger)\sim N(m^*(\thetavec^\dagger),v^*(\thetavec^\dagger))
\label{eq:likassump}
\end{equation}
where 
\begin{align*}
m^*(\thetavec^\dagger)&=m(\vec{\theta}^\dagger)+
K(\vec{\theta}^\dagger,\vec{\Theta})K(\vec{\Theta},\vec{\Theta})^{-1}\{\xvec(\vec{\Theta})-\vec{m}(\vec{\Theta})\}\\
v^*(\thetavec^\dagger)&=K(\vec{\theta}^\dagger,\vec{\theta}^\dagger)-K(\vec{\theta}^\dagger,\vec{\Theta})
K(\vec{\Theta},\vec{\Theta})^{-1}K(\vec{\theta}^\dagger,\vec{\Theta})^T.
\end{align*}
Strictly speaking, this emulator distribution is a function of the GP
hyperparameters and so we should average over their posterior
uncertainty. However, we have found that there is little gain in doing
this over using a delta approximation in which the hyperparameters are
fixed at their posterior mean.

We can incorporate the stochastic approximation of the GP emulator to
$\elogit~\widehat{p}_n$ into our observation model and integrate it out
to give a model with independent components of the form
\[
y\sim N\bigl(m^*(\thetavec),~
\sigma^2+v^*(\thetavec)+1/[n\,p^*_n(\thetavec)\{1-p^*_n(\thetavec)\}]\bigr)
\]
where
\[
p^*_n(\thetavec)=\text{eexpit}(\thetavec)=\{e^{m^*(\thetavec)}(1+0.5/n)-0.5/n\}/\{1+e^{m^*(\thetavec)}\}
\]
is the GP prediction at point~$\thetavec$ on the probability scale.

\section{History matching and validation}
\label{sec:historymatching}
It is quite likely that large parts of parameter space will give rise
to model outputs which are incompatible with the observed
proportions. In practice, we need our Gaussian processes to be
accurate over regions of parameter space with non-negligible posterior
support and so time spent fitting them in regions of negligible
posterior support is wasted. We can target the building of our GPs
within regions of non-negligible posterior support using methods known
as \textit{history matching} \citep{CummingG2010,Vernon2014}. These methods work
by determining which design points are implausible before undertaking
the computationally intensive task of evaluating the model output
(here empirical logits) at these points. After fitting a GP over,
implausible design points are determined by comparing data points
$y_i$ with their GP prediction. An implausibility measure at data
point $y_i$ takes the form
\[
I_i(\thetavec)=\frac{|y_i-m^*(\thetavec)|}
{\sqrt{\sigma^2+v^*(\thetavec)+1/[n\,p^*_n(\thetavec)\{1-p^*_n(\thetavec)\}]}}.
\]
Adopting a conservative strategy, the implausibility measure over the
entire data set is calculated as $I(\thetavec)=\max_i I_i(\thetavec)$.
Large values of $I(\thetavec)$ indicate that $\thetavec$ is an
implausible point in parameter space. It is common to declare points
as implausible if $I(\thetavec)>3$, this threshold being determined
using Pukelsheim's 3-sigma rule. Notice that the implausibility
measure requires a value to be given for the measurement error level.
We take the upper 1\% value of its prior distribution
($\sigma_D=0.58$) as this possibly overestimates $\sigma_D$ and in
doing so lowers the chance of incorrectly declaring a point as
implausible whilst still ruling out a significant proportion of
parameter space.

\subsection{Iterative fitting of Gaussian processes}
\begin{figure}
	\centering
	\includegraphics[width=\textwidth]{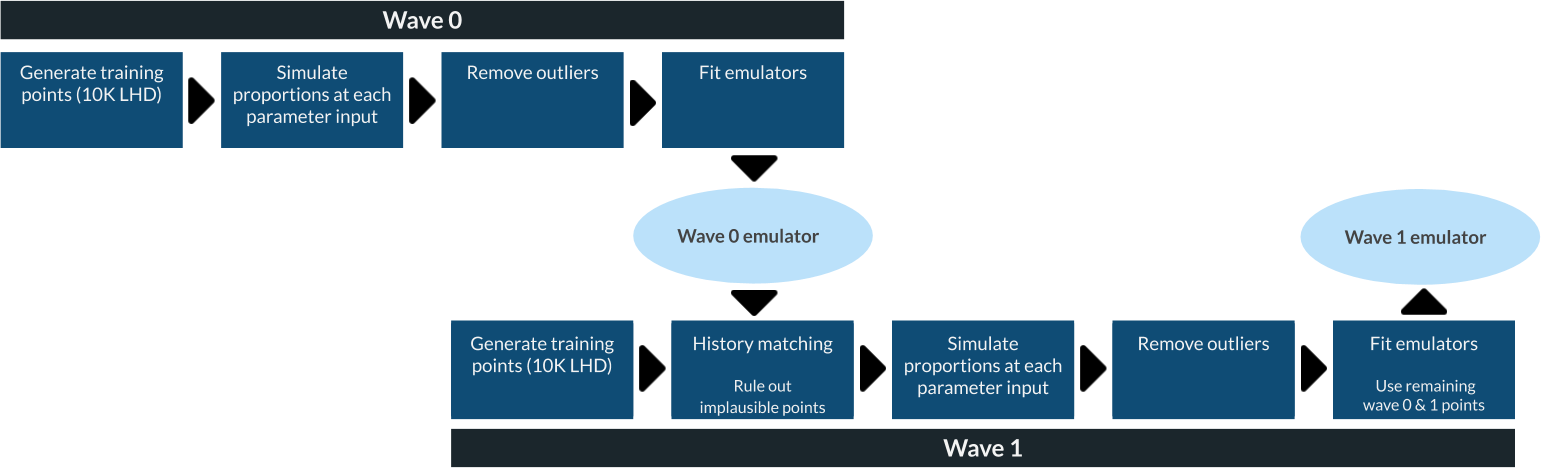}
	\caption{\ag{Flow chart showing the work flow involved in iteratively fitting Gausian process emulators.}}
	\label{fig:flow_chart}
\end{figure}

We used a sequential procedure for fitting the Gaussian process in
order to make best use of the limited computing resources available to
simulate realisations from the stochastic model \ag{(Figure \ref{fig:flow_chart})}. \ag{In what follows, 
we provide a brief description of this procedure and refer the 
reader to Appendix~C of the Supplementary Materials for further 
details.} 

We began in Wave~0 of
the fitting procedure by constructing a 10K-point LHD within the
cuboid defined by the 11 univariate marginal 99\% prior equi-tailed
intervals for each parameter~$\theta_j$. Multiple realisations
$(n=1000)$ were then obtained from the stochastic model at these
design points to determine the empirical logit proportions of cell
death and of inclusion bodies. Next we eliminated those design points
which gave proportions (of cell death or of inclusion bodies) outside
the range $(0.01,0.99)$, as such values look to be inconsistent with
the data and they take considerable computing time to obtain. Finally,
in Wave~0, we fitted a Gaussian process ($GP_0$) to the remaining 415
training points. Next, in Wave~1, we constructed an additional
10K-point LHD and removed any design points which had predicted
probabilities (of the form $\text{eexpit}\{m^*_{GP_0}(\thetavec_i)\}$)
outside $(0.01,0.99)$. The points surviving Wave~0 were then added and
any points with implausibilities $I_{GP_0}(\thetavec_i)>3$ removed.
Note that these implausibilities were calculated using the Gaussian
processes $GP_0$ fitted at the end of Wave~0. Multiple realisations
from the stochastic model were then simulated at the remaining 429
design points, empirical logits calculated and Gaussian processes
$GP_1$ fitted to this output. In practice, this process could be
repeated to give many more waves of history matching, and thereby
narrow down the plausible parameter space even further. However, due
to our limited computer resources, we terminated our process with
$GP_1$.

\subsection{Emulator validation}
\label{sec:validation}
Before using the $GP_{1}$ emulators as part of the inference scheme it
is important to verify that they provide an adequate description of
model realisations of the empirical logits. There are a variety of
diagnostic checks available based on comparisons of realisations from
the stochastic model at a new set of design points
($\thetavec^\dagger_i$) and comparing these with predictions made from
the emulators; see, for example, \cite{Bastos2009} and
\cite{Gneiting2007}.

\begin{figure}
\centering
\includegraphics[width=0.8\textwidth]{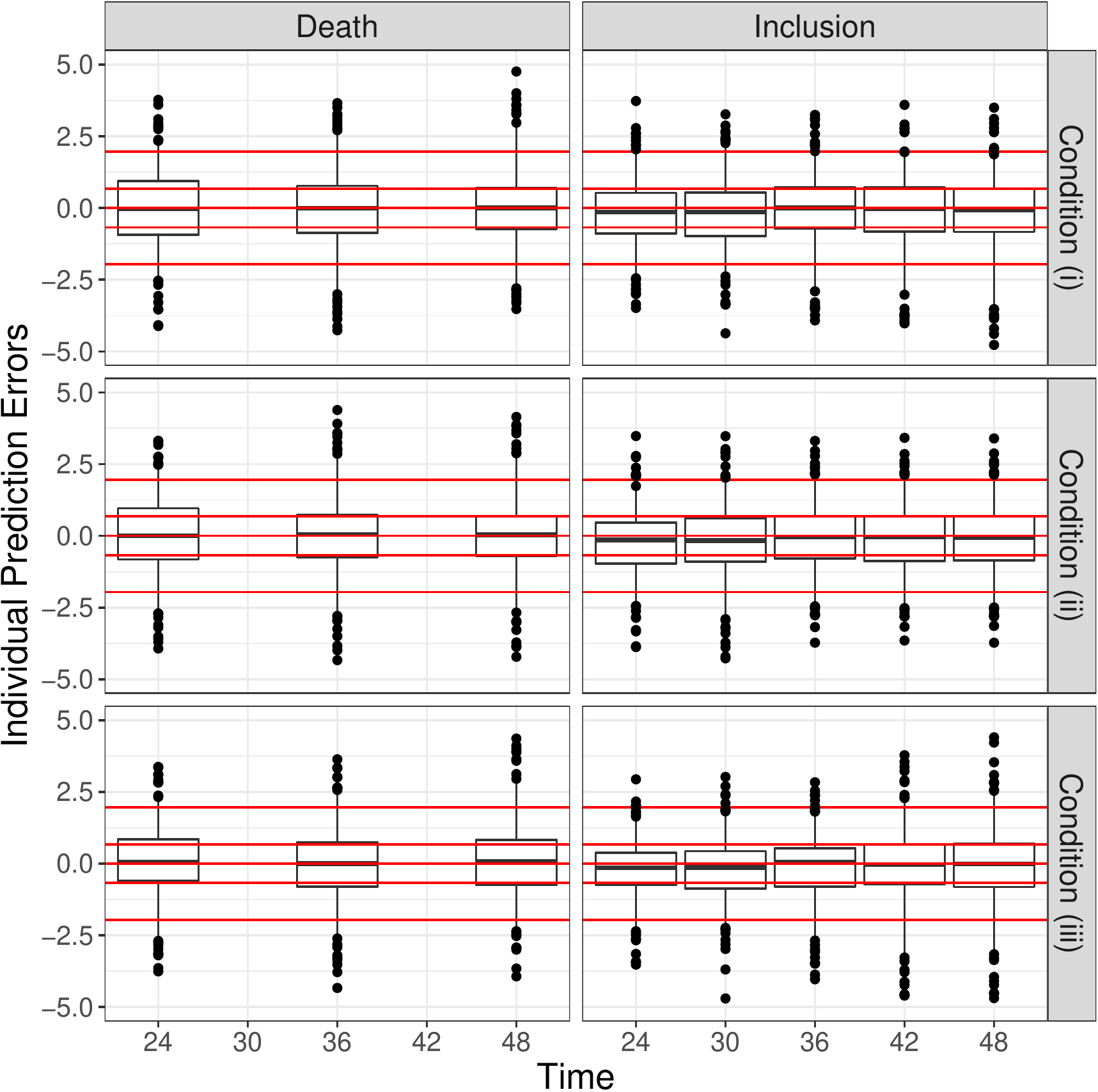}
\caption{Boxplots of the individual prediction errors for each
  emulator together with horizontal lines showing the position of the
  quartiles and upper and lower 2.5\% points of the standard normal
  distribution.}
\label{fig:polyq_diags}
\end{figure}

We constructed a validation design of a similar size to that used for
fitting the $GP_{\ag{1}}$ emulators by repeating the Wave~0--\ag{1}
procedure described in the previous section. This approach gives an
independent design but still contains points in regions of
non-negligible posterior support, where we most need the emulators to
fit well.  For each Gaussian process, we calculated individual
prediction errors (IPEs) at each design point as
$D(\vec{\theta}^{\dagger}_i) =
\{x(\vec{\theta}_i^{\dagger})-m^*(\vec{\theta_i}^{\dagger})\}/
\surd{v^*(\thetavec^{\dagger}_i)}$, where
$x(\vec{\theta}_i^{\dagger})$ is an empirical logit (for a particular
time-condition proportion) calculated from $n=1000$ runs of the
stochastic model with rates~$\thetavec=\thetavec_i^{\dagger}$.
Individual values of $D(\thetavec^\dagger_i)$ are informative about
emulator fit but it is particularly instructive to look at their
overall distribution. If the assumptions underpinning the emulator are
appropriate then the $D(\thetavec^\dagger_i)$ values should be a
random sample from a standard normal distribution. In particular, we
would expect that roughly $95\%$ of the IPE values are within the
interval $(-2,2)$.  If the magnitude of the IPE is too large, this
indicates that, at this input point, the emulator either fits poorly
or its variability is underestimated. Conversely, too many very small
values are indicative of an inflated emulator variance.
\cite{Gneiting2007} suggest that checks also be made using the
probability integral transform (PIT) to assess the (standard)
normality of the $D(\thetavec^\dagger_i)$ as, if this holds, values of
$\Phi\{D(\thetavec_i^\dagger)\}$ should form a random sample from a
standard uniform distribution, where $\Phi(\cdot)$ is the standard
normal distribution function. The IPEs for all 24 time-condition
emulators are shown in Figure~\ref{fig:polyq_diags}. Rather than
produce additional PIT plots we give boxplot summaries for each
emulator and include horizontal lines showing the positions of the
quartiles and upper and lower 2.5\% points of the standard normal
distribution.
These and other plots \ag{(see Appendix~D of the Supplementary Materials)} 
suggest that there are no strong departures from
the univariate normal assumptions \eqref{eq:likassump} used to form
the likelihood function.

\section{Posterior summaries and conclusions}
\label{sec:results}
\ag{To construct the emulators, training data (consisting of stochastic realisations from the model) 
were generated in parallel on a high performance computer (HPC), 
with 64 compute nodes, where each node had at least 128GB of RAM and 
a minimum processor speed of 2.5GHz. Hyper-parameter inference was performed 
via the RStan interface to Stan, on a different HPC (with 23 cores
and a minimum processor speed of 2.70GHz). The total CPU time to obtain 
the fitted emulators was approximately 1 day.} 

Realisations from the joint posterior distribution of all parameters of interest 
were obtained by using the fitted emulators and again implemented via the RStan interface to Stan. 

This MCMC algorithm uses the `no-U-Turn' Hamiltonian Monte Carlo sampler which efficiently
explores the parameter space to maximise mixing. We ran three chains, each for 100K
iterations, and using different initialisations. For each chain, the first half
(50K iterates) were discarded as warm-up/adaptation and the remaining half were
thinned to give 1K (almost) un-autocorrelated realisations from the posterior
distribution. Convergence was assessed by a variety of methods, including
graphical methods; see Appendix~E in the Supplementary Materials for example
traceplots and autocorrelation plots. We also examined diagnostics provided by
the \emph{shinystan} application~\citep{shinystan}. In particular, we looked at
the potential scale reduction statistic $\hat{R}$, which is the ratio of the
within-chain variance and between-chain variance of the posterior
sample~\citep{Gelman1992}. Here an $\hat{R}$-value close to one indicates that
all chains have reached equilibrium, and in our posterior sample, all parameters
had values less than~$1.1$. We also found that, for all parameters, the Monte
Carlo standard errors was less than $10\%$ of the posterior standard deviation
and the effective sample sizes greater than $10\%$ of the total sample size. 
\ag{Running the MCMC scheme for 100K iterations took approximately 7 days. 
Hence, the total computational cost (of obtaining the fitted emulators and 
performing the final calibration task) is of the order of 8 days. 
For comparison, consider use of the simulator directly inside an MCMC scheme. If we assume
that a single iteration takes 60 seconds, then 100K iteartions would take 70 days. We note
that this estimate is conservative, as in reality, the mixing of such a scheme may necessitate
many more iterations. Therefore, we expect that our use of emulators gives between 
one and two orders of magnitude increase in computational savings.}

\begin{figure}
\centering
\includegraphics[width=\textwidth]{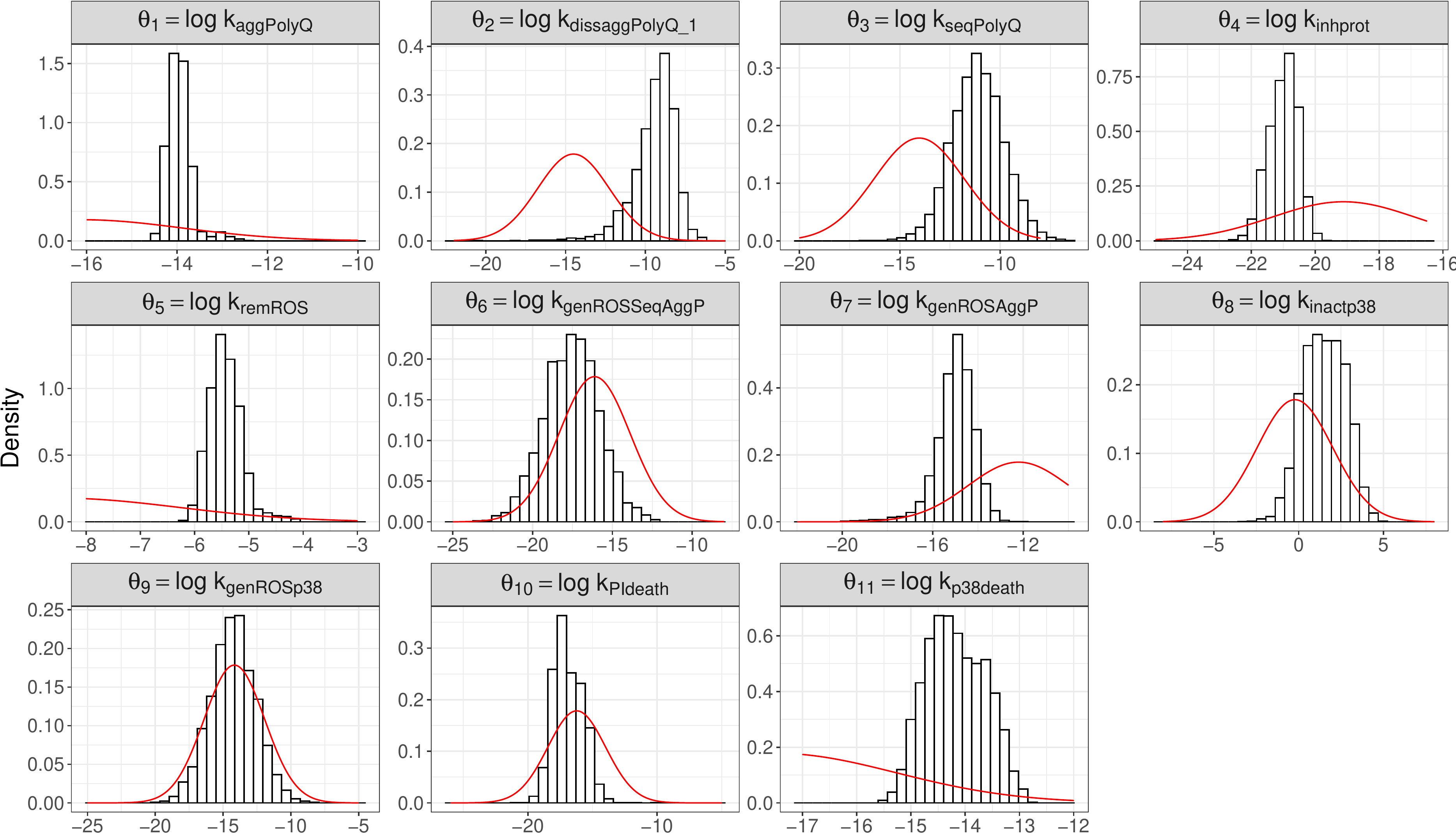}
\caption{Marginal prior and posterior distributions for the (logged)
  stochastic kinetic rate parameters (histogram -- posterior; red
  curve -- prior).}
\label{fig:post_kinetic}
\end{figure}

Graphical summaries of the marginal posterior distributions of the
(logged) kinetic rates are given in Figure~\ref{fig:post_kinetic}. The
figure shows that, despite the experimental data being thought to
perhaps give only a very partial insight into the biological
mechanism, it has in fact been very informative about many
parameters. 
The analysis has confirmed
values, with very much increased precision, for the rates of cell
death due to activation of p38 ($k_{p38death}$) and due to inhibition
of the proteasome by aggregates ($k_{PIdeath}$), which were also
observed but with no information on cause of death. It is of
particular note that the posterior distribution suggests that the
early stages of the aggregation process, when it is reversible,
probably occur at a much faster rate than had previously been
assumed. This suggests that both the disaggregation of small
aggregates and the formation and early growth of aggregates are more
rapid. Also the ratio of disaggregation and aggregation rates
$(k_{disaggPolyQ_1}/k_{aggPolyQ})$ is around an order of magnitude
higher than was thought. This suggests on the one hand that it may
take longer to reach the threshold size required for inclusion
formation but also that there will be more small aggregates present
which will inhibit the proteasome. \ag{The decrease in posterior mean also suggests that proteasome inhibition ($k_{inhprot}$) is likely to occur at
	a slower rate than first thought; probably due to originally underestimating the production of small aggregates.}

\ag{Looking at the rates involved in ROS turnover, the posterior suggests that the rate at which
ROS is removed (kremROS) is faster and that less ROS is generated by small aggregates (kgenROSAggP) than was thought, which means that the generation of ROS via the p38 pathway probably plays a larger contribution. This confirms the suggestion by Tang et al. (2010) that p38 inhibitors, an experimental intervention that they
tested, have therapeutic potential to reduce the detrimental effects of the aggregation process.}

\begin{figure}
\centering
\includegraphics[width=0.8\textwidth]{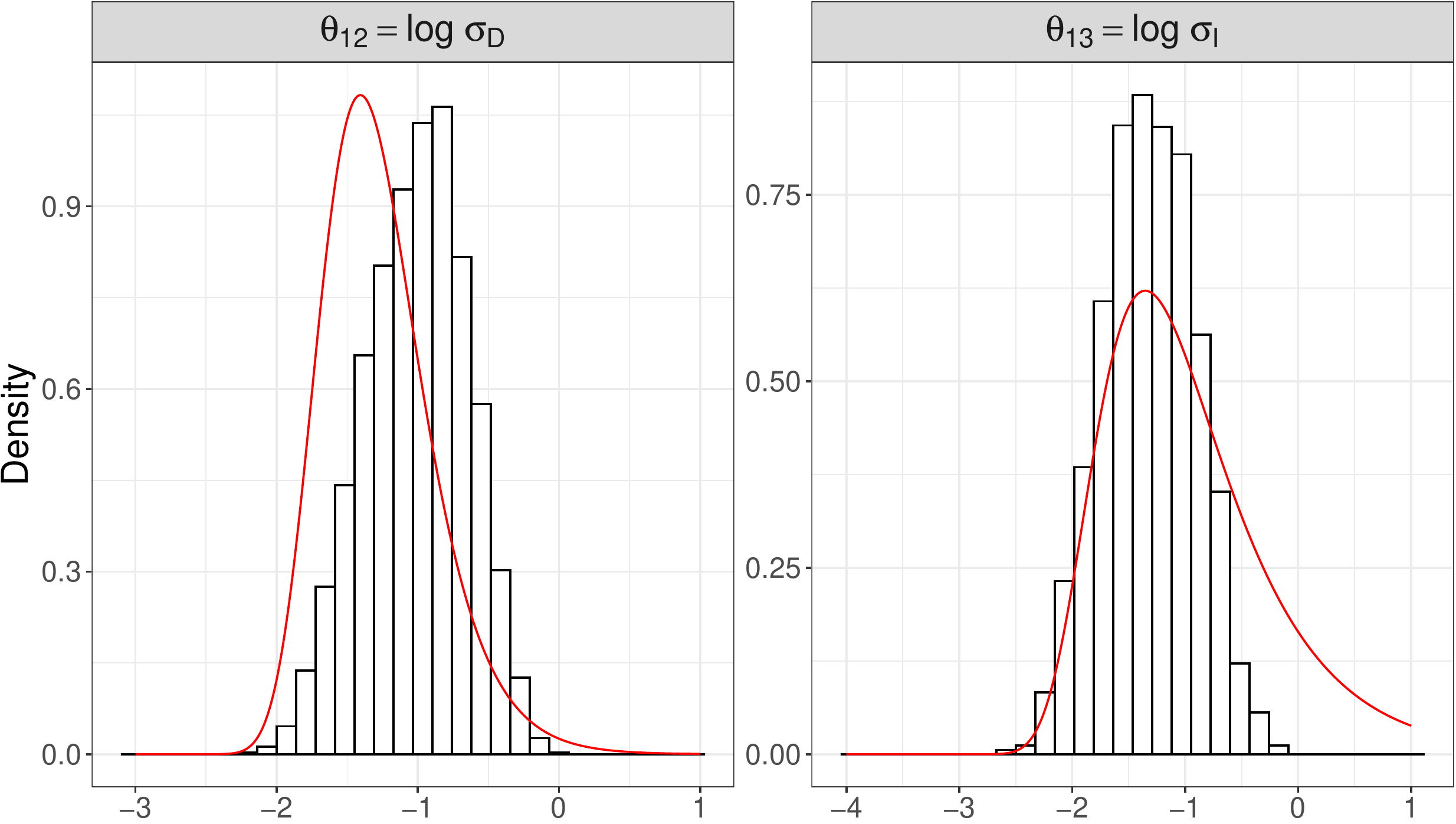}
\caption{Marginal posterior distributions for the (logged) measurement
  error parameters (histogram -- posterior; red curve -- prior).}
\label{fig:post_sigma}
\end{figure}

Figure~\ref{fig:post_sigma} shows summaries of the marginal posterior
distributions of the (logged) measurement error levels. 

\begin{figure}
\centering
\includegraphics[width=0.8\textwidth]{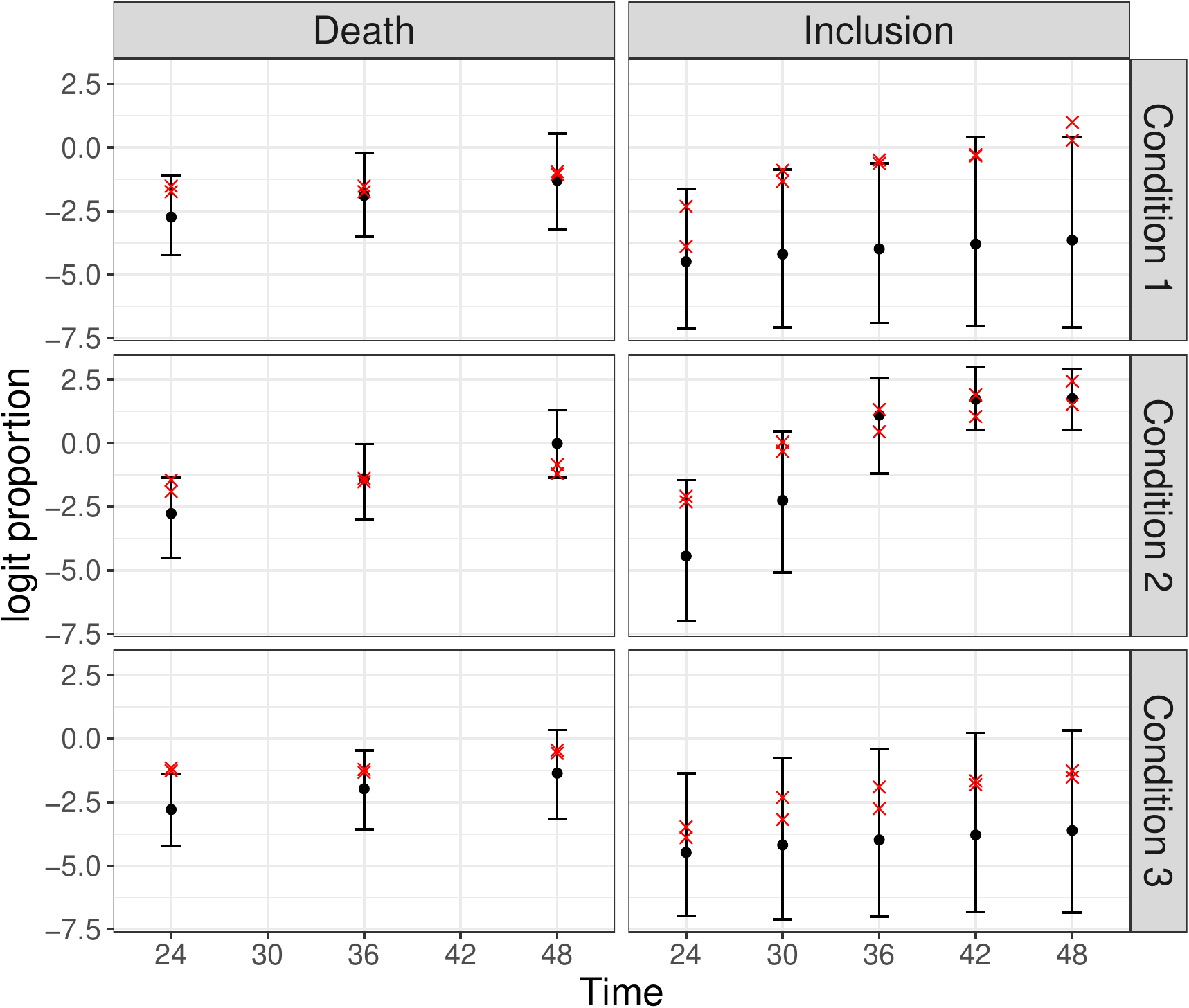}
\caption{Mean and 95$\%$ posterior predictive intervals for the
  proportions of cell death and of inclusions for each experimental
  condition, with the experimental data shown as crosses.}
\label{fig:predictive}
\end{figure}

We have also looked at the overall adequacy of the stochastic and observational
models by comparing the observed proportions under the three conditions with
their posterior predictive distribution. These distributions can be determined
by first running the stochastic model at a sample of realisations from the
posterior distribution of the model parameters and then obtaining realisations
from the observation model using this output. Figure~\ref{fig:predictive} shows,
for each experimental condition, 95\% predictive intervals for the proportions
of cell death or inclusion bodies on the logit scale, with the experimental data
shown as crosses. The positions of the data points in these intervals suggest
that there are no obvious departures from the joint observation-biological
stochastic model, though there is, of course, scope for model refinement and
further experiments. 


\section*{Acknowledgements}

This paper arose from work in the PhD project of Holly Fisher (n\'{e}e Ainsworth) who was funded by the Engineering and
Physical Sciences Research Council. We thank Doug Gray (University of Ottawa) for supplying us with the data and for
discussions leading to the calibration of our prior distribution.



\section*{Supplementary Materials}

Web Appendix 1 referenced in
Sections~\ref{sec:stochmodel}, \ref{sec:historymatching} and~\ref{sec:results}
is available upon request. The associated computer code is available from
\begin{center}
  https://github.com/csgillespie/polyglutamine-inference
\end{center}

\vspace*{-8pt}

\bibliographystyle{apalike} 
\bibliography{references.bib}

\end{document}